\documentclass[cup6a]{cupconf}
\RequirePackage{graphicx}%
\RequirePackage{epsf}%
\input{psfig.sty}

\title[Extragalactic metal-rich H\,{\small II} regions]{Measuring Chemical Abundances in Extragalactic Metal-Rich H\,{\Large II} Regions}

\author[F. Bresolin]{Fabio Bresolin}

\affiliation{Institute for Astronomy, University of Hawaii,  2680 Woodlawn Drive, 96822 USA}

\newcommand{\hii}{H\,{\sc ii}\rm}

\newcommand{\siii}{[S\,{\sc iii}]}
\newcommand{\nii}{[N\,{\sc ii}]}
\newcommand{\oiii}{[O\,{\sc iii}]}
\newcommand{\oii}{[O\,{\sc ii}]}
\newcommand{\sii}{[S\,{\sc ii}]}
\newcommand{\ariii}{[Ar\,{\sc iii}]}
\newcommand{\neii}{[Ne\,{\sc ii}]}
\newcommand{\neiii}{[Ne\,{\sc iii}]}

\newcommand{\lin}{$\,\lambda$}
\newcommand{\llin}{$\,\lambda\lambda$}

\newcommand{\apj}{ApJ}
\newcommand{\apjl}{ApJL}
\newcommand{\apjs}{ApJS}
\newcommand{\aap}{A\&A}
\newcommand{\mnras}{MNRAS}
\newcommand{\aj}{AJ}

\begin{document}
\maketitle

\begin{abstract}
The analysis of metal-rich \hii\/ regions has a profound impact on the calibration 
of abundance diagnostics widely used to measure the chemical content of 
star-forming galaxies, both locally and at high redshift. I review the main difficulties that affect direct abundance determinations from temperature-sensitive collisionally excited lines, and briefly discuss strong-line methods, in particular their empirical calibration. In the near future it will be possible to calibrate strong-line methods using metal recombination lines, providing abundances that are virtually insensitive to uncertainties on the nebular temperature structure.

\end{abstract}

\section{Chemical abundances of metal-rich H\,{\small II} regions: why?}

Ionized nebulae (\hii\/ regions) trace the sites of massive star formation in spiral and irregular galaxies. The rapid evolution of these stars, ending in supernovae explosions, and the subsequent recycling of nucleosynthesis products into the interstellar medium, make \hii\/ regions essential probes of the present-day chemical composition of star-forming galaxies across the Universe. The study of nebular abundances is therefore crucial for understanding the chemical evolution of galaxies. In the following pages I will provide an optical astronomer's perspective on some of the issues concerning the measurement of abundances in metal-rich \hii\/ regions, by focusing on the observational difficulties that are peculiar to the high metallicity regime, discussing some of the most recent abundance determinations from \hii\/ regions in the metal-rich zones of spiral galaxies, and indicating some possibilities for further progress.
Throughout this paper I will use the oxygen abundance as a proxy for the metallicity (oxygen makes up roughly half of the metal content of the interstellar medium), and assume the solar value from Asplund et al.~(2004), 12\,+\,log(O/H)$_\odot$\,=\,8.66. Elements besides oxygen will not be discussed in great detail.

\subsection{Motivations}
Why measure abundances of metal-rich \hii\/ regions? After all, as we will see in Section~2, metal-rich \hii\/ regions pose difficulties to the observer that are not present at lower metallicities, i.e.~roughly below half the solar O/H value.
However, high abundances are encountered in a variety of astrophysical contexts, and the study of ionized nebulae often provides the only way to measure these abundances. Here are a few examples that motivate detailed studies of metal-rich \hii\/ regions, aimed at improving our abundance diagnostics:

\begin{enumerate}
\renewcommand{\theenumi}{\alph{enumi}}

\item galactic abundance gradients have been known in spiral galaxies since the pioneering work carried out in the 1970's (see the compilations by Vila-Costas \& Edmunds~1992, Zaritsky et al.~1994). Even after the recent downward revision of the metallicity in metal-rich \hii\/ regions from empirical methods (Bresolin et al.~2004), the central parts of spiral galaxies reach or exceed the solar metallicity (Pilyugin et al.~2004, 2006a).

\item chemical evolution models of galaxies aim at explaining the observed abundance gradients and abundance ratios on the basis of assumptions about the stellar yields and initial mass function, the star formation rate and efficiency, the presence of gas inflows and outflows, etc. (Henry et al.~2000, Chiappini et al.~2003, Carigi et al.~2005). A major uncertainty in the input data regards the metallicity of the inner regions of spiral galaxies.

\item the luminosity-metallicity relation of local ($z<0.1$) star-forming galaxies depends critically on the methods adopted for measuring abundances in nearby \hii\/ regions, in particular at the high-metallicity end (Salzer et al.~2005). A similar uncertainty exists for more remote galaxies ($z<0.9$, Kobulnicky \& Kewley~2004, Maier et al.~2005). The finding that the brightest (and more massive) galaxies exceed the solar metallicity by a factor of a few [up to 12\,+\,log(O/H)\,=\,9.2] is dependent on the calibration of strong-line methods, such as those discussed in Section~3.

\item the same conclusion can be drawn when considering the mass-metallicity relation for low- and high-redshift ($z\leq3$) star-forming galaxies (Tremonti et al.~2004, Savaglio et al.~2005, Erb et al.~2006).

\item solar-like metallicities are measured (via strong-line methods) at the largest redshifts so far explored (Shapley et al.~2004; see M.~Pettini's contribution in this volume). Once more, the resulting abundances and the implications for the chemical evolution of early galaxies rely on the accuracy of local calibrations of strong-line methods.

\item the number ratio of Wolf-Rayet stars of type WC to those of type WN is predicted to be an increasing function of the stellar metallicity (Meynet \& Maeder~2005; P.~Crowther, this volume). We have observational confirmations of this prediction (Massey 2003, Crockett et al.~2006), although the results in the high-metallicity regime are somewhat fuzzy due to uncertainties in measuring abundances for the Wolf-Rayet parent galaxies.

\item despite claims that the upper limit of the stellar initial mass function 
of the ionizing clusters of \hii\/ regions might
decrease with increasing abundance (Thornley et al.~2000), other studies  suggest that this is not the case (Pindao et al.~2002). Obtaining accurate abundances at the metal-rich end is critical for the reliability of this conclusion.

\item a metallicity dependence of the Cepheid period-luminosity relation is possible, although the issue is far from being settled. Sakai et al.~(2004) showed how the distances derived to metal-rich galaxies are sensitive on the method adopted to measure the metallicity  from \hii\/ regions.

\end{enumerate}

\section{Difficulties at high metallicity}

Obtaining a {\em direct} measurement of chemical abundances in \hii\/ regions requires that we have a good idea of the value of the electron temperature, $T_e$. The strength of the forbidden emission lines (collisionally excited) from various metal ions commonly detected in nebular spectra are strongly sensitive to $T_e$, besides to the abundance of the originating ionic species, $N(X^{+i})$: 
\vspace{-0.5cm}

\begin{equation}
I_\lambda \propto N(X^{+i})~ \epsilon_\lambda
\end{equation}
\vspace{-0.3cm}

\noindent
since the line emissivity $\epsilon_\lambda$ has essentially an exponential dependence on $T_e$:

\vspace{-0.2cm}
\begin{equation}
\epsilon_\lambda \propto \Omega(T)\,T_e^{-0.5}\,e^{-\frac{\chi}{k\,T_e}}.
\end{equation}
\vspace{-0.3cm}

\noindent
Taking the ratio of forbidden lines corresponding to atomic transitions that originate from widely separated energy levels, such that their relative population is sensitive to the electron temperature, provides a measure of $T_e$. One uses the ratio of the {\em auroral} lines (transition from the second lowest excited level to the lowest excited level) to the corresponding {\em nebular} lines (first excited level to the ground level). The auroral lines are generally quite faint, and relatively high temperatures are required to excite enough atoms via collisions to the energy levels from which these lines originate.
The case of oxygen is probably the most well-known; specifically, for \oiii\/ one uses the ratio of \lin4363 (auroral) to \llin4959,\,5007 (nebular; the ground level is actually split by spin-spin interactions, hence the two nebular lines). 

The heart of the problem in the high-metallicity nebular business  is illustrated in Fig.~1. An increase in the metallicity has the effect of increasing the cooling in the nebula, which occurs mostly via line emission from metals, mostly oxygen,
first through the \oiii\/ optical forbidden lines, and at higher abundances (lower temperatures) through the far-IR \oiii\/ hyperfine transitions at\llin52,\,88\,$\mu$m. Fig.~1 shows that as one moves to low temperatures (say $T_e < 8000\,K$) the \oiii\,4363/(4959+5007) line ratio becomes soon very small, due to its exponential dependence. In fact, for extragalactic work the ratio hits the typical limit imposed by the capabilities of current 8-to-10m telescopes (represented by the horizontally streched rectangle in Fig.~1) already for metallicities well below the solar value (notice the position of the solar symbol on the dotted curve, around $6000\,K$).

Qualitatively the picture remains the same when considering auroral-to-nebular line ratios for other ions, however we can consider cases where these ratios
are larger, at a given temperature, than in the case of \oiii. This happens 
for ions where the upper level (from which the auroral transition originates) lies at a lower energy above the ground with respect to the \oiii\lin4363 case (5.3 eV), for example for \nii\lin5755 (4.0 eV) and \siii\lin6312 (3.4 eV). Observationally this situation creates a big advantage: even at super-solar metallicities ratios such as \siii\,6312/(9069+9532) or \nii\,5755/(6548+6583) are measureable in bright extragalactic \hii\/ regions with current instrumentation.
The solar symbols on the \siii\/ and \nii\/ curves, representing the typical line ratios measured for solar metallicity extragalactic \hii\/ regions (e.g. Bresolin et al.~2005) lie well above the typical observational limit of 8-10m telescopes. Note that some of the coolest extragalactic nebulae for which a direct $T_e$ determination exists have temperatures in the range between 5000\,$K$ and 6000\,$K$.

\begin{figure}
\centerline{\psfig{file=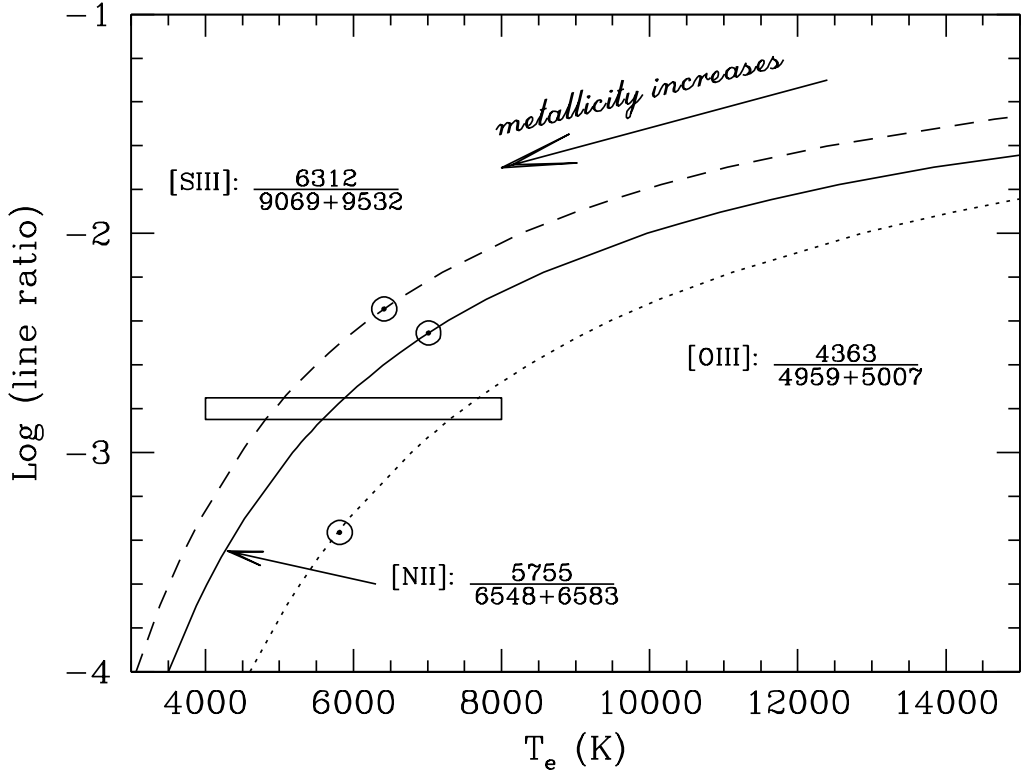,width=\textwidth}}
\caption{Dependence of commonly used auroral-to-nebular line ratios on electron temperature $T_e$. The arrow points in the direction of increasing metallicity (increased cooling results in lower $T_e$). In the high-metallicity regime, $T_e$ is typically below 8000\,$K$. The solar symbols on each of the three curves indicate
the line ratios observed for extragalactic \hii\/ regions of approximately solar
metallicity. The open rectangle represents the current observational limit reached with an 8m-class telescope. While it is hopeless to measure electron temperatures of metal-rich nebulae from observations of the \oiii\lin4363 auroral line, it is possible to derive $T_e$ (and therefore abundances) using the auroral lines \siii\lin6312 and \nii\lin5755 even at metallicities above solar.
}
\label{auroral}
\end{figure}

\subsection{Caveat 1: assumptions  on the temperature structure of ionized nebulae}

The analysis of the chemical composition of extragalactic \hii\/ regions based on collisionally excited lines commonly simplifies the structure of the nebulae as composed by spherically symmetric, concentric shells, centered on the ionizing source. Different ionic species dominate
the emission in different shells. Typically 
the \oiii\/ and \neiii\/  lines are assumed to originate
in a central, high-excitation zone, while the \oii, \nii\/ and \sii\/ lines are emitted in the low-excitation outer region. An intermediate-excitation zone is sometimes introduced for \siii\/ and \ariii. Each of these emitting zones is characterized by different ionic temperatures, e.g. T(O$^{++}$), T(O$^{+}$), and so forth, which can differ from the line temperatures measured from the auroral-to-nebular line ratios, such as T\oiii, T\oii, etc.

Photoionization models provide useful relationships between the temperatures in the various zones (Stasi\'nska~1978). For example, Garnett (1992) found: 
\begin{equation}
\rm T[S\,{\scriptstyle III}]\,=\,0.83\,T[O\,{\scriptstyle III}]\,+\,1700~\it K 
\end{equation}

\vspace{-0.4cm}

\begin{equation}
\rm T[O\,{\scriptstyle II}]\,=\,T[N\,{\scriptstyle II}]\,=\,T[S\,{\scriptstyle II}]\,=\,0.70\,T[O\,{\scriptstyle III}]\,+\,3000~\it K.
\end{equation}

\medskip
\noindent
Similar relations have been found by other authors (Campbell et al.~1986, Pilyugin et al.~2006b).
Since at high metallicity [approximately 12\,+\,log(O/H)\,$>$\,8.5] the \oiii\/ temperature cannot be measured directly using the \lin4363 auroral line, one can then calculate T\oiii\/ from either T\siii\/ or T\nii, or some clever combination of the available temperatures. It is important to remember that virtually all \oiii\/ ionic abundances measured in high-metallicity extragalactic \hii\/ regions are derived using this technique.
Real ionized nebulae have naturally a much more complicated structure than assumed in this stratification model (see the study of NGC~588 in M33 by Jamet et al.~2005). Some reassurance on the validity of the assumed temperature relations for real nebulae comes from the works of Kennicutt et al.~(2003) on M101 and Bresolin et al.~(2004) on M51.

\section{Strong-line methods}
Often we cannot rely on the direct method described in Sect.~2 to measure nebular abundances. This occurs whenever auroral lines cannot be detected, for example in the case of the spectrum of a cool \hii\/ region of very high metallicity, or in the case of the spectrum of a faint, high-redshift galaxy, where only the brightest emission lines can be seen. 
Among the various strong-line methods, the $R_{23}$ indicator originally proposed by Pagel et al.~(1979) stands out as arguably the most popular. For in-depth discussions on this and additional methods, which will be discussed only marginally  here, the reader should consult the papers by Kewley \& Dopita (2002) and by P\'erez-Montero \& D\'{\i}az (2005). 

In the Pagel et al.~method, one considers the two most important stages of ionization of oxygen, both emitting collisionally excited lines in the visually accessible range of the spectrum: $R_{23}$\,=\,(\oii\lin3727\,+\,\oiii\llin4959,\,5007)/H$\beta$. Like all strong-line diagnostics,
$R_{23}$ as an abundance indicator has a statistical value, based on the fact that the hardness of the ionizing radiation correlates with  metallicity. Considering \hii\/ regions of increasing metallicity (lower temperature), the optical \oiii\/ lines become progressively fainter, as the emission shifts to the far-IR fine-structure lines. 

Calibrations of $R_{23}$ in terms of the nebular chemical composition abund in the literature. Many are based on grids of photoionization models (Edmunds \& Pagel~1984, McGaugh~1991, Kewley \& Dopita~2002). Empirical calibrations are based on abundances derived from auroral lines (Pilyugin~2001). The end-user of any of these calibrations should always keep in mind the systematic abundance differences that one obtains using different calibrations, up to 0.5 dex, from the same input emission line fluxes.

\subsection{Empirical calibrations from deep spectroscopy of extragalactic H\,{\small II} regions}

Necessary input data for empirical calibrations of strong-line methods are obviously the direct abundances obtained from detailed studies of high-metallicity \hii\/ regions, focusing on the inner zones of giant spiral galaxies. Early works on the abundances of high-metallicity \hii\/ regions from 
auroral lines include those of Kinkel \& Rosa (1994) on the region Searle~5 in M101 (anchor point at high metallicity of the early Pagel et al.~1979 calibration), Castellanos et al.~(2002) on CDT1 in NGC~1232, and Kennicutt et al.~(2003) on Searle~5 and H1013 in M101. The availability of 8m-class telescopes in recent years has expanded the field considerably. About a dozen \hii\/ regions have now direct abundance determinations in M51 (Bresolin et al.~2004, Garnett et al.~2004), and a larger sample has been studied by Bresolin et al.~(2005) in five southern spirals.

Figure~2 shows the relation between $R_{23}$ and the nebular abundance derived from the availability of auroral lines, essentially \siii\lin6312 and \nii\lin5755, as described earlier. The data points have been taken from the literature, emphasizing with larger symbols the extension to low $R_{23}$ values (higher abundance) derived from observations with 8m-class telescopes.
The double-valued nature of the $R_{23}$ indicator is apparent; the degeneracy can be removed through the availability of emission line diagnostics that are monotonic with the oxygen abundance (e.g.~\nii\lin6583/H$\alpha$). The scatter in the data shows the sensitivity of $R_{23}$ to additional parameters besides the chemical abundance, such as the ionization parameter and the effective temperature of the ionizing stars (P\'erez-Montero \& D\'{\i}az~2005).
Note that nebular abundances still hover around the solar value even at the smallest observed $R_{23}$, and only a handful of points appear to be more metal-rich than the solar value. In addition, the curves in Fig.~2 clearly illustrate the well-known discrepancy between $R_{23}$ calibrations obtained empirically (here taken from the P-method of Pilyugin \& Thuan~2005) from auroral line observations and those obtained from photoionization model grids (here Kewley \& Dopita~2002 is taken as a representative example). This discrepancy amounts to approximately 0.3 dex, and its complete explanation still eludes us.

\subsection{Implications of the recent empirical results}
The direct abundances obtained recently in M51 and in other metal-rich galactic environments have important consequences for the derivation of chemical abundances from strong-line methods. These results in fact concern most of the investigations mentioned in the Introduction. For example, when looking at the radial abundance gradients in spiral galaxies we obtain a decrease in the central O/H values, by up to a factor of three, relative to the results based on older $R_{23}$ calibrations, or in general those based on photoionization model grids (Kennicutt et al.~2003). The abundance gradients become consequently flatter.
The central abundances of spirals regarded to be among the most metal-rich in our neighborhood (such as M51, NGC~3351, M83) are basically solar, or exceed the solar value by just a small amount, lying in the range 12\,+\,log(O/H)\,=\,8.60--8.75 (Pilyugin et al.~2006a). Values up to three times solar had been claimed earlier for some of these galaxies.
It must be pointed out that when considering the luminosity-metallicity and the mass-metallicity relationships for star-forming galaxies the adoption of different calibrations has the effect of modifying the zero points, but the general trends remain essentially unaltered (Ellison \& Kewley~2005).

\begin{figure}
\centerline{\psfig{file=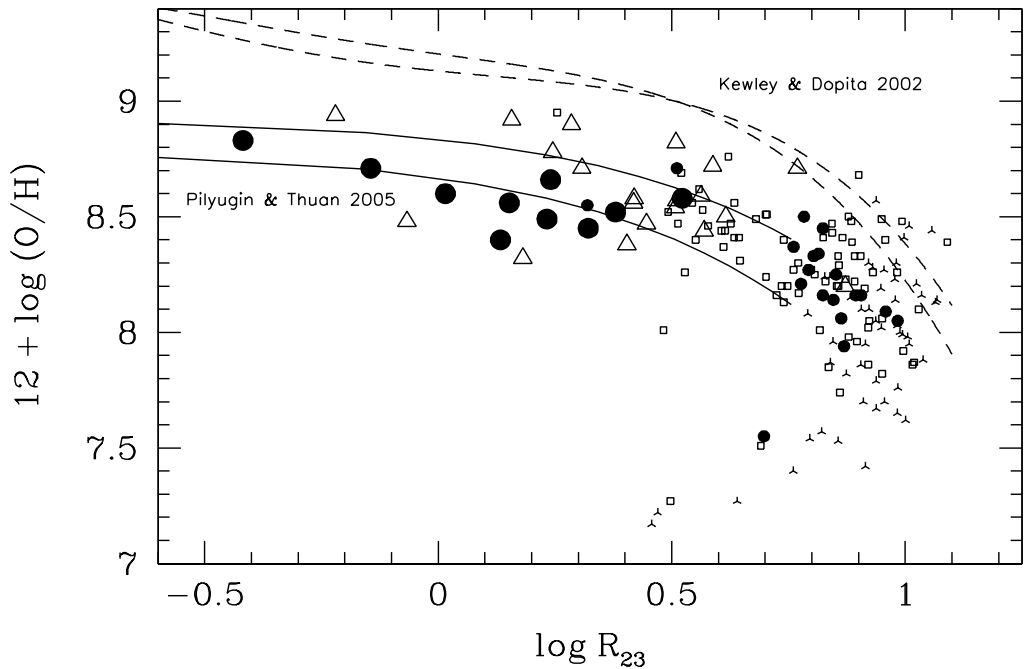,width=\textwidth}}
\caption{This abundance $vs.$~$R_{23}$ diagram shows observational data for extragalactic \hii\/ regions, for which the O/H abundances have been derived from the detection of auroral lines (incomplete collection drawn from various sources in the literature). The high-metallicity samples of Bresolin et al.~(2004, 2005) are shown with enlarged symbols (full circles and triangles). The small full circles are from the M101 sample studied by Kennicutt et al.~(2003).
The curves show the calibrations of the $R_{23}$ indicator derived empirically by Pilyugin \& Thuan~(2005) and from photoionization model grids by Kewley \& Dopita (2002). In both cases two curves are drawn, corresponding to two different values of the ionization parameter.
}
\label{r23}
\end{figure}

\subsection{Other strong-line methods}
It should be at least briefly mentioned here that there are other strong-line methods beside $R_{23}$ that can be used to determine abundances of high-metallicity star-forming regions. They rely on emission lines that can generally be measured even in low signal-to-ratio nebular spectra, such as \nii\lin6583 and \sii\llin6717,\,6731. The accuracy of the derived O/H measurements should always be regarded to be at the 0.2-0.3 dex level.
Among the most useful methods for the analysis of high-redshift star-forming galaxies is N2\,=\,\nii\lin6583/H$\alpha$ (Denicol\'{o} et al.~2002, Pettini \& Pagel~2004), because the emission lines involved are very close in wavelength, and can be observed with the same setup when moving for example to the infrared $K$ band, where the rest-frame spectral lines end up at redshift $z\sim2$.
Among the drawbacks affecting N2 are the saturation above the solar abundance and the sensitivity to the chemical history of nitrogen.

The sulphur line-based $S_{23}$\,=\,(\sii\llin6717,\,6731\,+\,\siii\llin9069,\,9532)/H$\beta$ (V\'{\i}lchez \& Esteban~1996)
has been calibrated in terms of both oxygen and sulphur abundances (P\'erez-Montero et al.~2006). The advantage of this indicator derives from its small sensitivity to the ionization parameter compared to $R_{23}$, but its usefulness is limited to redshifts $z<0.1$. Recently Stasi\'nska (2006) has introduced abundance indicators based on high-excitation lines only: \ariii\lin7135/\oiii\lin5007 and \siii\lin9069/\oiii\lin5007, that alleviate one potential problem of indicators involving low-excitation lines, such as N2, namely the fact that the latter lines can be produced both in the diffuse galactic interstellar medium, as well as in the star-forming regions.

\subsection{Caveat 2: temperature gradients}
Metal-rich \hii\/ regions are not isothermal clouds of ionized gas: both large-scale temperature gradients and small-scale fluctuations are predicted to exist.
The presence of large-scale $T_e$ gradients is a consequence of the efficient cooling of the inner regions via far-IR \oiii\/ fine-structure lines (there are no corresponding fine-structure \oii\/ lines). Strong gradients are expected to develop above the solar metallicity, with the inner region cooler by several thousands of degrees with respect to the outer zones (Stasi\'nska~1980, Garnett~1992). This has important consequences for the chemical analysis. Due to the exponential dependence of the line emissivities on $T_e$, the observed line strengths are weighted strongly toward warmer regions. At high metallicity the optical \oiii\/ collisionally excited lines originate therefore not in the cooler, high-excitation zone, but instead in the warmer O$^+$ zone. As a consequence, using the auroral-to-nebular line ratio scheme to derive $T_e$ leads to over-estimate the ionic temperature: T\oiii\,$>$\,T(O$^{++}$). Under these conditions, the oxygen abundance is under-estimated, possibly by large factors (Stasi\'nska~2005).

\section{Seeking independence from the temperature dependence: metal recombination lines}

How reliable are the chemical abundances derived from the auroral line method at high metallicity? Are the effects of temperature gradients measureable? There are a number of ways to carry out these tests. Using objects other than \hii\/ regions to trace galactic abundance gradients is  becoming feasible with high signal-to-noise spectroscopy of planetary nebulae and luminous stars in nearby galaxies.
As an example, the blue supergiant metallicities derived by Urbaneja et al.~(2005) in M33 are in good agreement with the \hii\/ region abundances by V\'{\i}lchez et al.~(1988) and Crockett et al.~(2006).
Detailed photoionization modeling also offers a possible venue, however, as mentioned earlier, current discrepancies between empirical and model-based abundances are not fully understood. A further approach envisions the analysis of nebular lines that depend far less than the collisionally excited lines on the electron temperature, such as optical metal recombination lines and IR fine-structure lines. In both  cases the line emissivity is only moderately dependent on $T_e$, even though it can be a strong function of the gas density. For example, considering Equation (2.2) for fine-structure transitions, in which the excitation potentials are very small (in the range 0.01-0.04 eV  $vs.$ 2-3 eV for optical transitions), the exponential term becomes virtually unit, and only a moderate $T_e$ dependence is left. 

The use of \oiii\llin52,\,88\,$\mu$m to derive abundances in extragalactic \hii\/ regions is limited to only a few cases in the literature. Garnett et al.~(2004) have used ISO spectra of region CCM10 in M51 to derive 12\,+\,log(O/H)\,=\,8.8, which compares to 12\,+\,log(O/H)\,=\,8.6 obtained by Bresolin et al.~(2004) on the basis of auroral line analysis. 
Additional far-IR studies include those of \hii\/ regions in the Milky Way by Peeters et al.~(2002), Mart\'{\i}n-Hern\'andez et al.~(2002) and Rudolph et al.~(2006).
Alternatively, it is also possible to explore the mid-IR, accessible to Spitzer, to study the \neii\lin12.8\,$\mu$m and \neiii\lin15.6\,$\mu$m lines, as done in M33 by Willner \& Nelson-Patel~(2002).

The emissivity of recombination lines from metals is also only weakly dependent 
on $T_e$: $\epsilon_\lambda \sim \alpha_{\!\mbox{\tiny\em eff}}(\lambda) \sim T_e^{-1}$. When measuring abundances relative to hydrogen, the $T_e$ dependence becomes negligible, since H recombination lines have a similar sensitivity to $T_e$.
The main drawback in their use for abundance studies is their weakness. Among the strongest metal recombination lines are the {\sc o\,ii} multiplet at 4651\,\AA, and {\sc c\,ii}\lin4267. These lines are beautifully detected in the high-resolution spectra of some well-known \hii\/ regions in the Milky Way, such as the Orion nebula (Esteban et al.~2004) or the Trifid nebula (Garc\'{\i}a-Rojas et al.~2006). Detection in extragalactic nebulae can still be challenging, and results for only one relatively high-metallicity [12\,+\,log(O/H)\,=\,8.5] \hii\/ region, NGC5461 in M101 (Esteban et al.~2002) have been published so far.

In general, it is found that oxygen abundances obtained from recombination lines are 0.2-0.3 dex larger than those derived from collisionally excited lines under the assumption (tacitly implied in the discussion so far) that temperature fluctuations are absent in \hii\/ regions. A similar result is obtained when comparing O/H derived from a combination of optical and far-IR \oiii\/ lines or from optical data alone (Jamet et al.~2005).
This systematic difference can be explained with the existence of temperature fluctuations in \hii\/ regions, as argued by M.~Peimbert and collaborators in a series of papers (see Peimbert et al.~2006 for a recent review). With 
mean square temperature fluctuations $t^2$ in the 0.02--0.06 range, one can reconcile the abundances from collisionally excited lines with those from recombination lines for the \hii\/ regions where both sets of emission lines have been analyzed.

Figure 3 shows the same diagram of Figure~2, with the addition of Galactic and extragalactic \hii\/ regions for which a comparison between abundances obtained from collisionally excited lines and from metal recombination lines has been carried out so far (sources can be found in Peimbert \& Peimbert~2005 and Peimbert et al.~2006). These objects populate the diagram in its upper branch only down to log\,$R_{23}\simeq0.5$.
Efforts are under way to extend this sample  to higher metallicities with high-resolution spectroscopy at 8-to-10m telescopes. 
Recently, Bresolin (2006, in prep.) has measured the {\sc c\,ii}\lin4267 line, as well as the mean square temperature fluctuation $t^2\,=\,0.06$, in H1013, an \hii\/ region in the inner, metal-rich parts of M101. The oxygen abundance derived within the scheme developed originally by Peimbert (1967) and Peimbert \& Costero (1969) is 
12\,+\,log(O/H)\,=\,8.9 (1.7\,$\times$ solar), about 0.3 dex larger than the value obtained not accounting for temperature fluctuations. These two values are represented by the open and solid square symbols in Fig.~3.

The results obtained from recombination lines are in good agreement with the $R_{23}$ calibration derived from photoionization models, as Figure~3 shows. This is a consequence of the fact that, even though photoionization models fail at reproducing the intensity of an auroral line like \oiii\lin4363, calling for an as yet unidentified heating source (Stasi\'nska \& Schaerer~1999), the predictions for the strong lines \oii\lin3727 and \oiii\llin4959,\,5007 are less sensitive to the temperature structure of the nebulae. The virtual independence of the abundances obtained from metal recombination lines on $T_e$ makes them more suitable to provide an accurate calibration of strong-line methods, such as $R_{23}$, than collisionally excited lines. The extension of the recombination line measurements to the high-metallicity regime becomes then imperative.

\begin{figure}
\centerline{\psfig{file=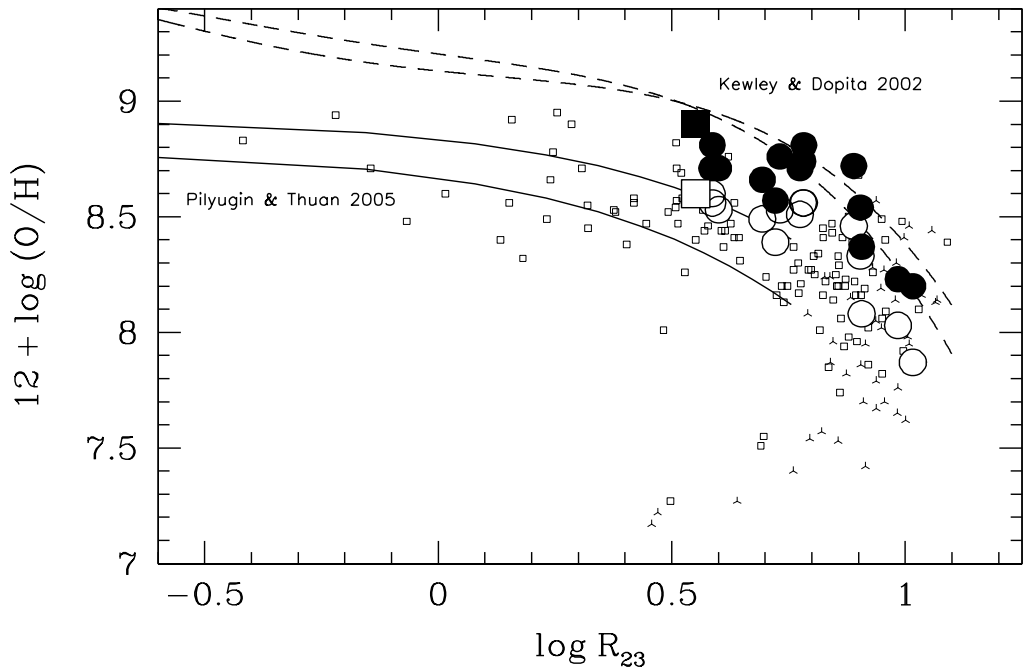,width=\textwidth}}
\caption{Same as Figure~2, with the observational data points shown with small symbols. The larger symbols represent Galactic and extragalactic \hii\/ regions for which abundances have been derived both from collisionally excited lines (open symbols) and from metal recombination lines (full symbols; sources in Peimbert et al.~2006). The \hii\/ region H1013 in M101 studied by Bresolin (2006) is shown by squares. }
\label{r23_2}
\end{figure}

\end{document}